\documentclass[12pt]{iopart}
\usepackage{supertabular,vmargin}
\usepackage[latin1]{inputenc}
\usepackage[T1]{fontenc}
\usepackage{graphicx}
\usepackage{stmaryrd}
\usepackage{ae,aeguill,hyperref,multicol}
\usepackage[LGR,T1]{fontenc}
\usepackage{subfigure}
\usepackage{index}
\usepackage{array}
\usepackage{color}
\DeclareGraphicsExtensions{.jpg,.png,.bmp,.tiff,.eps}
\begin{document}

\title[Collapse of antiferromagnetism in CeRh$_{2}$Si$_{2}$ : volume versus entropy]{Collapse of antiferromagnetism in CeRh$_{2}$Si$_{2}$  : volume versus entropy}

\author{A. Villaume $^1$, D. Aoki $^1$, Y. Haga $^3$, G. Knebel $^1$, R. Boursier $^1$ $^{2}$, J. Flouquet $^1$}

\address{$^1$ CEA-Grenoble, DSM/DRFMC/SPSMS 38054 Grenoble Cedex-9, France}

\address{$^2$ Laboratoire des Champs Magnétiques intenses, CNRS, 38045 Grenoble Cedex-9, France}

\address{$^3$ Advanced Science Research Center, Japan Atomic Agency, Ibaraki 319-1195, Tokai, Japan}

\begin{abstract}
The thermal expansion of the heavy fermion compound CeRh$_2$Si$_2$ has been measured under pressure as a function of temperature using strain gages. A large anomaly associated to the Néel temperature has been detected even above the suspected critical pressure $P_c \sim$ 1.05 GPa where no indication of antiferromagnetism has been observed in calorimetry experiments sensitive to the entropy change. An unexpected feature is the pressure slowdown of the antiferromagnetic-paramagnetic transition by comparison to the fast pressure collapse predicted for homogeneous first order quantum phase transition with one unique pressure singularity at $P_c$. A large pressure dependance is observed in the anisotropy of the thermal expansion measured parallel or perpendicular to the c axis of this tetragonal crystal. The Fermi surface reconstruction associated to the first order transition produces quite different pressure response in the transport scattering measured along different crystallographic directions. A brief discussion is made on other examples of first order quantum transitions in strongly correlated electronic systems : MnSi and CeCoIn$_5$.

\end{abstract}




\section{Introduction}

\indent The disappearance of the long range magnetic ordering as a function of an external parameter like pressure or magnetic field at a critical point in strongly correlated electron systems (SCES) is still an open question \cite{Lohn:07}. Associated changes of Fermi surface (FS) or the role of valence or charge instabilities close to the critical point are under debate. Even the nature (second order or first order) of the quantum phase transition at zero temperature from antiferromagnetic (AF) to paramagnetic (PM) ground states is questioned. Experimental as well as theoretical supports lead to the conclusion that a ferromagnetic quantum critical point will be first order \cite{Bel:05}.

Even in heavy fermion compounds (HFC) which are well suitable systems for such a study, very few measurements concern a careful study in the AF state on approaching the critical pressure $P_{c}$ where AF will switch to PM ground state. Most discussions focus on non-Fermi liquid behavior observed in the PM state at finite temperature as P tends to $P_{c}$ \cite{Lohn:07, Flou:06}. This deficiency in the study of the collapse of the antiferromagnetism from the AF side is due to the fact that the signals related to the collapse of the order parameter (the sublattice magnetization $M_{0}(T))$ becomes very weak in neutron scattering as well as in calorimetry or transport measurements close to P$_{c}$ \cite{Flou:06}.

By contrast to these probes, the relative volume (V) variation or its temperature derivative (the thermal expansion $\alpha$) is expected to be easier to measure. For a second order phase transition, the pressure variation of the Néel temperature of an AF is expected to be large near $P_{c}$ leading even to a divergence of its Grüneisen parameter ($\Omega_{T_{N}}=\frac{-\partial Log T_{N}}{\partial Log V}$) in the framework of the spin fluctuation theory \cite{Mor:03, Kamb:97, Zhu:03}. For a first order transition, a volume discontinuity must occur at $P_{c}$ since the Nernst principle requires that the entropy reaches zero at $T\rightarrow0K$ whatever is the order of the transition. Two decades ago, it was already pointed out that HFC are characterized at very low temperature by a large electronic Grüneisen parameter $\Omega^{*}=-\frac{\partial{LogT^{*}}}{\partial{LogV}}$ where $k_BT^{*}$ stands for the dominant low energy scale which can be the Kondo temperature, the spin fluctuation or valence fluctuation temperature \cite{Benoit:81, Takke:1981}. 

Here we concentrate on the HFC CeRh$_2$Si$_2$. At ambiant pressure, CeRh$_2$Si$_2$ presents two magnetic transition : the first transition is second order at ambiant pressure, with $T_{N1}\sim 35 K$ while the second one is first order at ambiant pressure, with $T_{N2}\sim 25 K$. CeRh$_2$Si$_2$ is an ideal system to study the quantum phase transition from AF to PM as the Néel temperature $T_{N1}\sim 35 K$ at ambiant pressure is high, but the critical pressure $P_c\sim 1 GPa$ is rather low. The first transition at $T_{N1}$ is suspected to become first order above 1.0 GPa since pronounced FS reconstruction has been observed \cite{Ara:02}.

In this article we present detailed measurements of the length variation of the heavy fermion compound CeRh$_2$Si$_2$ as a function of pressure. Some comparisons with ac calorimetry and inelastic neutron diffusion measurements are given, and the evolution of the nature of the magnetic transition at $T_{N1}$ under pressure is discussed.

\section{Experimental details}

Single crystal of CeRh$_2$Si$_2$ were grown by the Czochralski pulling method. A sample of 2.5x2.5x3mm size was cut from a large crystal. To measure the relative elongation under pressure, a 2x2.5mm Kyowa$^{\textregistered}$ strain gage was first glued along the c axis direction of the sample using the Kyowa$^{\textregistered}$ UC-27A glue and following a well established Kyowa procedure \cite{Kyow:07}. The $\Delta{R}$ change of the initial strain gage's resistance R is related to the relative elongation $\varepsilon$ by \ref{eq_strain1} :
\begin{equation} \label{eq_strain1}
\frac{\Delta{R}}{R}=K_s.\varepsilon
\end{equation}
R stands for the gage resistance at 300 K without any strain (R$\simeq$120$\Omega$) and $K_s$ is the gage factor. which is near 2 in general purpose strain gages. The temperature variation of $K_s$ is known and does not exceed 2\% between 2 K and 300 K. A measurement along the a axis direction was also performed : each relative elongation can either be measured separately or in opposition.

A $^{4}$He cryostat (1.6 K to 300 K) enables us to perform low temperature measurements. Since the resistance of the electric wires varies between the inside and the outside of the cryostat, we use a thermo-compensation systems which consists of a strain gage glued on a tungsten carbide (CW) reference sample and a Wheatsone electrical bridge. The bridge is balanced at low temperature (30 K) by a precision adjustable potentiometer. Using a lock-in detection, a precision higher than 5x10$^{-7}$ for the relative elongation of CeRh$_2$Si$_2$ is achieved.

Measurements under pressure up to 1.5 GPa can be performed inside a home made CuBe pressure cell. The CW reference and the CeRh$_2$Si$_2$ sample are put inside a teflon cartridge which is filled with Daphne$^{\textregistered}$ oil. This cartridge is put inside the pressure cell. A precise determination of the pressure value inside the cell is achieved at ambiant temperature by a calibrated manganin sensor. At low temperature the pressure is determined by the superconducting transition of lead \cite{Wittig:66}. The precision on the value of the pressure value inside the cell is less than 0.005 GPa. Previously calorimetic measurements had been realized with the ac technique now extensively used in Grenoble \cite{Derr:01}.

\section{Experimental results}

\begin{figure}[h]
\begin{center}
\includegraphics[height=0.60\linewidth, angle=270]{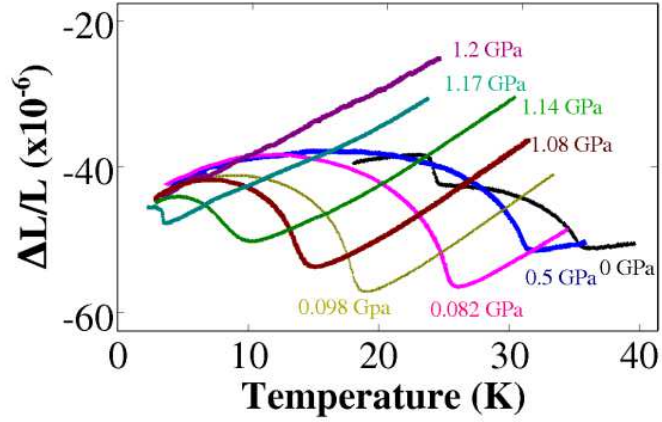}
\caption{{\small Relative variation $\Delta L/L$ along the c-axis as a function of the temperature for different pressures}}\label{dilat} 
\end{center}
\end{figure}

Figure \ref{dilat} shows the relative length variation $\Delta$L/L along the c-axis for several pressures. At P=0, on cooling the first AF transition occurs at $T_{N1}$ = 35 K; a second transition appears in the AF state at $T_{N2} \sim$ 24 K. This second transition disappears above P $\sim$ 0.5 GPa \cite{Kawa:00}. In this article, we focus on the disappearance of AF order i.e. on the pressure variation of $T_{N1}$ and related low temperature properties.

\begin{figure}[h]
\begin{center}
\includegraphics[height=0.40\linewidth]{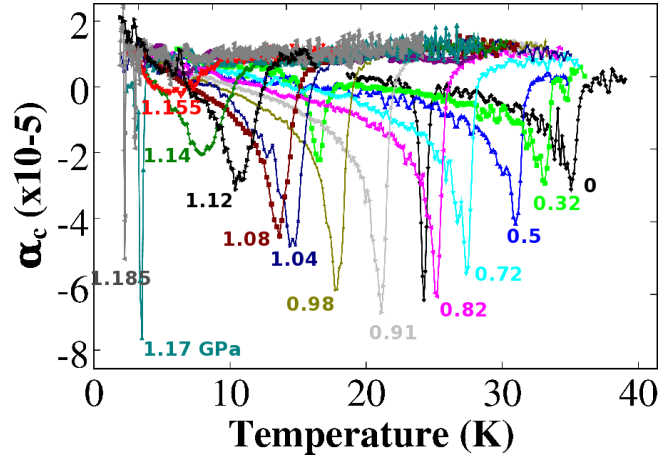}
\caption{{\small Thermal expansion coefficient $\alpha_{c}$ as a function of the temperature for different pressures in GPa}}\label{coeff_dilat} 
\end{center}
\end{figure}

It is worthwhile to notice that up to 1.0 GPa the signal does not change drastically. However at 1.14 GPa it becomes weaker and surprisingly at 1.17 GPa a real discontinuity in $\Delta L/L$ occurs. The data at the 15 different pressures are drawn on figure \ref{coeff_dilat} where the thermal expansion coefficient along the c-axis ($\alpha_{c} = \frac{\partial}{\partial T}(\Delta{L}/L)$) is represented. The measured thermal expansion takes into account the relative elongation of the CW reference sample (about 1.1x10$^{-6}$ K$^{-1}$). $\alpha_{c}$ obtained by dilatation measurements at P=0 GPa is 10\% lower than $\alpha_{c}$ reported in previous capacitive measurements \cite{Ara:98}. 

One can see clearly the assymetric signal observed at $T_{N}$ below 1.0 GPa and suddenly at 1.17 and 1.18GPa a real discontinuity in $\Delta{L}/L$. At low pressure the change in $\alpha_c$ is associated with the P variation of the specific heat and the P increase in the slope of $\frac{\partial T_{N}}{\partial P}$, according to the Ehrenfest relation for a second order phase transition , $\frac{\partial{P}}{\partial{T}}=\frac{\Delta{C}}{T.v.\Delta\alpha}$, where $C$ is the thermal capacity, $T$ the temperature, $v$ the molar volume and $\alpha$ the thermal expansion coefficient.

\begin{figure}[h]
\begin{center}
\includegraphics[height=0.60\linewidth, angle=270]{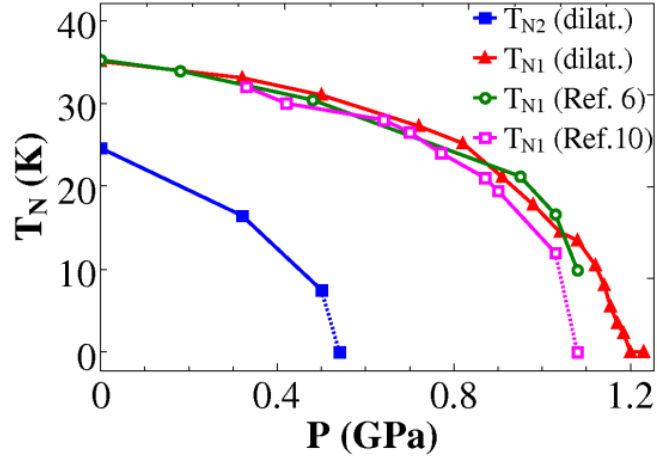}
\caption{{\small Phase diagram of CeRh$_2$Si$_2$ obtained by dilatation measurements, neutron scaterring \cite{Kawa:00} and ac calorimetry. \cite{Ara:02}}}\label{diag_phase} 
\end{center}
\end{figure}

\begin{figure}[h]
\begin{center}
\includegraphics[height=0.45\linewidth]{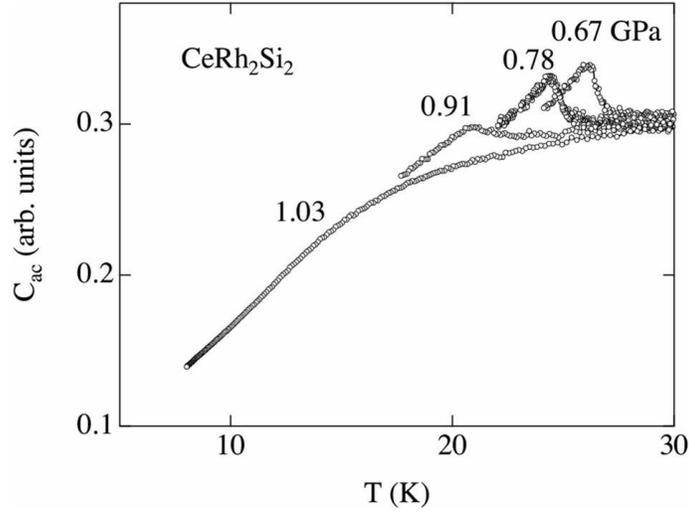}
\caption{{\small ac specific heat of CeRh$_2$Si$_2$ with the broadening and collapse of the AF anomaly on approaching the critical pressure $P_c$.}}\label{Haga} 
\end{center}
\end{figure}

Figure \ref{diag_phase} represents the pressure depandance of $T_{N}$ which is defined by the minima of $\frac{\partial}{\partial T}(\Delta{L}/L)$.
Rather good agreement is found between $T_N$ determined by different techniques at zero pressure. But also under pressure at least when a signal associated to $T_N$ can be detected in resistivity $T_N$(P), ac calorimetry $T_N$(C), inelastic neutron scattering $T_N(M_0$) or dilatation experiments, a good agreement is found in the determination of $T_N$. It is worthwhile to notice that a clear entrance in the AF ordering is observed at $T_{N1}$ in this experiment above 1.04 GPa while no signal can be detected in resistivity experiments above 1.03 GPa \cite{Ara:02, Oha:03, Gro:97} and also a very broadened signal can be seen in calorimetry experiments only up to $P \sim 1.03$ GPa as shown on figure \ref{Haga}.

\begin{figure}[h]
\begin{center}
\includegraphics[height=0.60\linewidth, angle=270]{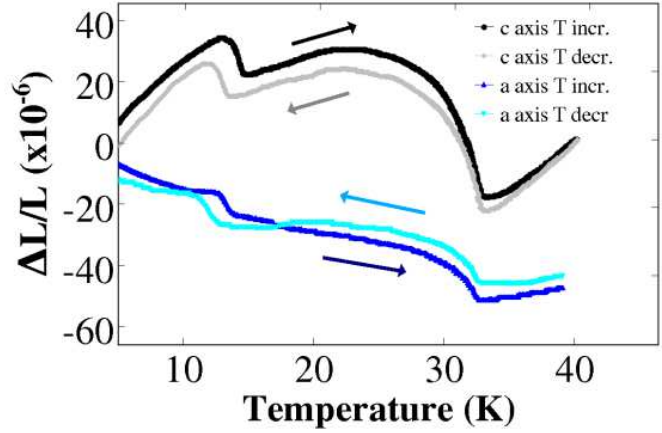}
\caption{{\small Relative variation $\Delta L/L$ of along the c-axis and the a-axis at 0.41GPa.}}\label{compa_ac} 
\end{center}
\end{figure}

The pressure variation of the ratio $\frac{\alpha_a}{\alpha_c}$ has been also determinated. Large pressure dependence of $\frac{\alpha_a}{\alpha_c}$ at T$_{N1}$ is observed : $\frac{\alpha_a}{\alpha_c}$ increases continuously from 0.23, 0.73 and 1 respectively for P=0.41 GPa, 0.77 and 1.05 GPa. An example of the $\frac{\Delta{L}}{L}$ curves along a and c axis is given on figure \ref{compa_ac} for P=0.41 GPa. An interesting point is the existence of an hysteresis in the transition associated with $T_{N2}$. Such a behavior has already been reported by Ohashi et al \cite{Oha:03} from resistivity measurements. This hysteresis is found only near $T_{N2}$ for both c and a axis, confirming the first order nature of the transition. No hysteresis is detected at $T_{N1}$ in agreement with the assumption that below 1GPa the transition is second order (excellent agreement with the Ehrenfest relation).

\begin{figure}[h]
\begin{center}
\includegraphics[height=0.40\linewidth]{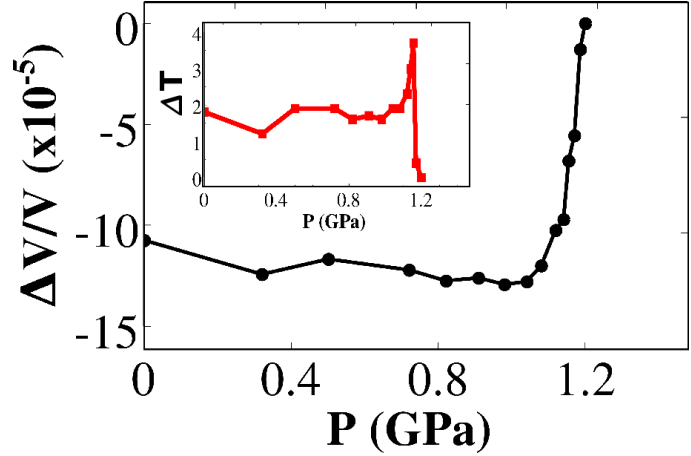}
\caption{{\small Relative change of volume at the AF transition extracted from figure \ref{coeff_dilat}. Inset : temperature broadening of $T_{N1}$ (middle height width calculated on the thermal expansion coefficient of the figure \ref{coeff_dilat}).}}\label{change_V_1} 
\end{center}
\end{figure}

The volume change associated to the AF transition is represented in figure \ref{change_V_1} by supposing an average P value of $\alpha_a=\frac{\alpha_c}{2}$. $\frac{\Delta{V}}{V}$ is calculated by substracting the CW reference contribution and by integrating the area below each curve represented in figure \ref{coeff_dilat}.

A clear change of regime appears for $P \sim$ 1.05 GPa. Roughly the same characteristic pressure is detected in the broadening ($\Delta$T) of $\alpha_c$ at $T_{N1}$ where $\Delta T$ stands for the full width half maximum of the $\alpha_c$ versus T curves (insert of figure \ref{change_V_1}). 

The two surprising results are that 
\begin{enumerate}
 \item at the opposite to both expected previsions of AF spin fluctuation ($T_{N1} \sim (P_c-P)^{2/3}$, \cite{Flou:06}) and first order transition, $T_{N1}$ appears to extrapolate linearly with P to zero down to $P_c\sim$ 1.2 GPa and not with an infinite slope.
 \item a real volume discontinuity is detected in the vicinity of $P_c$ which points to a first order transition.
\end{enumerate}

\begin{figure}[h!]
\begin{center}
\includegraphics[height=0.40\linewidth]{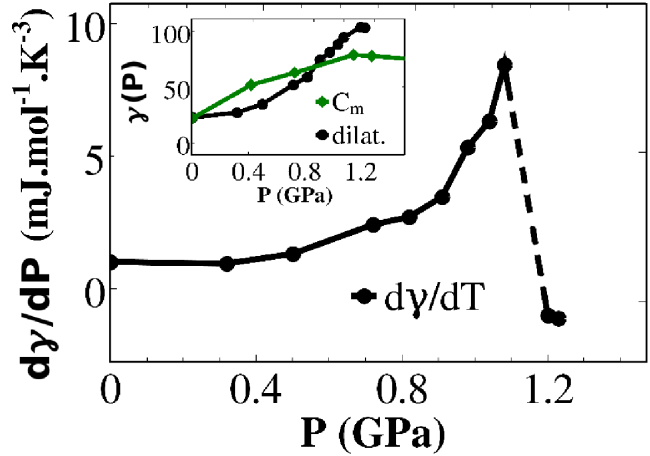}
\caption{{\small Evaluation of $\frac{\partial\gamma}{\partial{P}}$(P) and of $\gamma$(P) in mJ.mol$^{-1}$.K$^{-2}$ (insert) by dilatation measurements. Comparison with specific heat (inset) \cite{Graf:97}.}}\label{gamma} 
\end{center}
\end{figure}

Up to now only the regime close to $T_{N1}$ was analyzed. Below $T_{N1}$, the low temperature behavior is dominated by the P variation of the average effective mass $m^{*}$ which is related to the pressure derivative of the entropy (S) according to the Maxwell relation \ref{Maxwell} : 
\begin{equation} \label{Maxwell}
\frac{\partial{V}}{\partial{T}}=-\frac{\partial{S}}{\partial{P}} \hspace*{1cm} 
\end{equation}
The pressure derivative variation of the Sommerfeld coefficient of the specific heat $\gamma$ ($\frac{\partial{\gamma}}{{\partial{P}}}\propto-\frac{1}{T}\frac{\partial{V}}{\partial{T}}$) is shown in figure \ref{gamma}. For $P>1.08 GPa$, $T_{N1}$ is too close from our low temperature limit and the determination of $\frac{\partial{\gamma}}{\partial{P}}$ is not possible. The insert also shows the pressure variation of $\gamma$ determinated by previous calorimetric measurements \cite{Graf:97} and now here by dilatation.

As indicated in figure \ref{gamma}, the direct specific heat measurements suggest a maxima of $\gamma$ at $P_c$ equal to 80 mJ.mol$^{-1}$K$^{-2}$ \cite{Graf:97}. This value is not too far from our extrapolation regarding the crudeness in the pressure variation of the $\frac{\alpha_a}{\alpha_c}$ ratio and also a different hydrostaticity between our pressure cell (with at room temperature liquid transmitting medium) and the previous specific heat measurements (with solid transmitting medium). The relative pressure variation of $\gamma$ obtained by integration of $\frac{\partial\gamma}{\partial{P}}$ is in good agreement with effects observed in HFC \cite{Flou:06} and the relative pressure variation of the A coefficient of the resistivity described below.The maximum of $\gamma$ and of $\Delta{T}$ may not correspond to the collapse of $T_{N1}$ which occurs for $P \sim 1.2$ GPa as it is predicted for a second order magnetic quantum critical point in spin fluctuation theory. We will analyse later the possible origin of such a shift.

\section{Discussion}

Strain gage probes are confirmed to be an unique tool to detect the magnetic ordering close to its collapse via a first order transition. Only few direct accurate measurements of the volume variation under pressure close to quantum criticality have been performed. The relative accuracy of the relative volume change $\frac{\Delta{V}}{V}$ determinated using strain gages is better than 5x10$^{-7}$, i.e. one order of magnitude by comparison to X-ray and two order of magnitude higher by respect to neutron scattering. The power of strain gages was recently reemphasized by the discovery of the first order quantum transition between the hidden order phase and the antiferromagnetic phase of URu$_2$Si$_2$ as function of pressure \cite{Mot:03}.

As pointed out above, the resistivity anomaly associated to $T_{N1}$ becomes small close to $P_c$. An indirect way to detect the enhancement of $m^{*}$ and the mark of a complex behavior close to $P_c$ is to look in details to the low temperature variation of the resistivity generally described by \ref{ro(T)} :

\begin{equation} \label{ro(T)}
\rho(T,P) = \rho_{0}(P) + A_{n}(P)T^{n}
\end{equation}

In difference to many HFC close to $P_c$, in CeRh$_2$Si$_2$ a clear Fermi liquid regime is detected (n=2) on both sides of $P_c$, and even close to $P_c$ \cite{Ara:02, Oha:03, Gro:97}. The persistence of this FL behavior even up to rather high temperature ($T\simeq 1 K$) is quite unique for Ce HFC where usually close to $P_c$ a non FL behavior in resistivity (n<2) persists well below 1K in the vicinity of $P_c$. This singularity is caused by the strong first order nature of the transition from AF to PM. Our experiment indicates a relative volume discontinuity $\frac{\Delta{V}}{V}$ at $P_c$ of at least 10$^{-5}$. By comparison to a well known example such as the first order transition or the melting curve of solid $^{3}$He where $\frac{\Delta{V}}{V} \simeq 5\%$ but $P_c$ only 34 bars \cite{Flou:06}, in CeRh$_2$Si$_2$, $\frac{\Delta{V}}{V}$ is reduced roughly by three order of magnitude but $P_c$ is increased by roughly two order of magnitude. The mechanical work P$\Delta$V at $P_c$ in CeRh$_2$Si$_2$ is only 10\% smaller than the one furnished in $^{3}$He. Thus the first order nature of the transition in CeRh$_2$Si$_2$ above 1.0GPa is not weak.

\begin{figure}[h!]
\begin{center}
\includegraphics[height=0.50\linewidth]{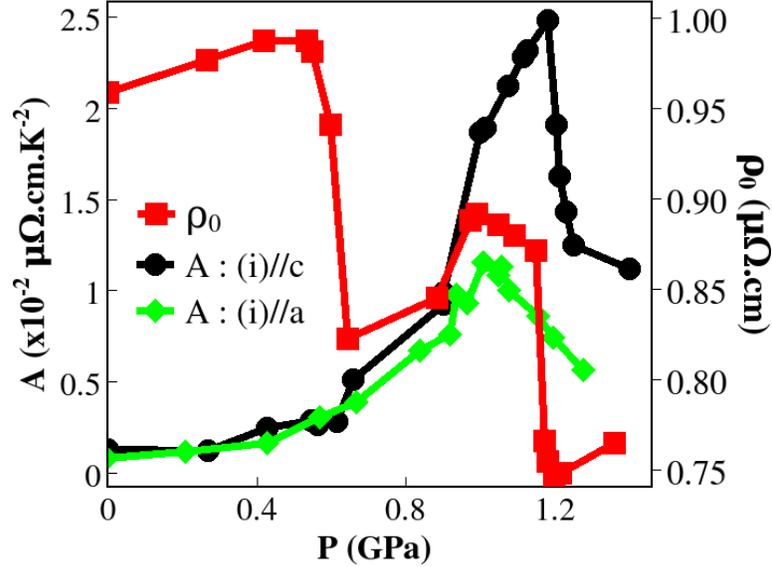}
\caption{{\small Pressure variation of the inelastic term A for the current (i) applied parallel to a axis \cite{Bou:06} and parallel to c axis \cite{Ara:02}.}}\label{A_rodolphe} 
\end{center}
\end{figure}

A recent careful analysis \cite{Bou:06} of the electrical resistivity obtained with the current i applied in the basal plane has confirmed the existence of large P changes in residual resistivity and in the $T^{2}$ term \cite{Oha:03} not only around a given maxima $P_c$ but for a pressure range from 1.03 to 1.2 GPa, as shown in figure \ref{A_rodolphe}. This is in good agreement with a pressure slot found in our thermal expansion experiments.

Careful studies of the resistivity around $P_c$ were reported previously with the current along the c axis \cite{Ara:02}. No such a singular behavior was found within a similar pressure interval. The pressure variation of A and $\rho_0$ near $P_c$ is quite stronger for the current applied along the a axis than along the c axis. This difference is related to the pressure variation of the magnetic anisotropy detected in susceptibility measurements : a collapse of the anisotropy in the magnetic susceptibility under pressure measured parallel and perpendicular to the c axis is observed at $T_{N1}$ for $P\rightarrow P_c$ \cite{Mori:00}. This tendency of an isotropic magnetic behavior is directly related to the isotropy observed in the thermal variation of $\alpha_a$ and $\alpha_c$ at low temperature near 1 GPa. This converging behavior points out the quasi-coincidence between the magnetic quantum critical point ($P_c$) and the valence critical (or crossover) pressure ($P_v$) where the 4f electron will loose its sensitivity to the crystal fiel effect \cite{Flou:06}. By contrast, for the CeRu$_2$Si$_2$ familly, a strong decoupling occurs between $P_c \sim -0.3 GPa$ and $P_v \sim 4 GPa$ and a quasi constant value of $\frac{\alpha_a}{\alpha_c} \sim 0.3$ is observed on both side of $P_c$ for pressure sweeps near 1 GPa \cite{Lacer:89}. In CeRh$_2$Si$_2$, $P_c$ and $P_v$ are not discernible. Our guess will be that $\frac{\alpha_a}{\alpha_c} \rightarrow 1$ for $P>P_v$. Thus the pressure evolution of the ratio $\frac{\alpha_a}{\alpha_c}$ is a new interesting feature rarely considered under pressure.

The damping of the Ising type anisotropy under pressure is caused by the fast pressure increase of the Kondo temperature $T_K$. At P=0, the estimation of $T_K$ is between 30 K-100 K \cite{Sev:89, Kawa:98}, and the crystal field splitting $C_{CF}$ is near 300 K \cite{Sev2:89}. Already near $P_c$, the Kondo effect energy reaches $C_{CF}$. As for CeIn$_3$ or CePd$_2$Si$_2$, the magnetic instability at $P_c$ coincides or is very close to the pressure $P_v$ where the Ce centres enter in their intermediate valent regime \cite{Flou:06}. With a Kondo Grüneisen parameter $\Omega^{*}(T_K)\simeq 100$, $T_K$ will double at $P_c$ while $C_{CF}$ is suspected to be weakly pressure dependent, thus $k_BT_K$ will rapidly overcome $C_{CF}$ above $P_c$.

The striking difference in the behavior of the scattering detected by resistivity for the current applied parallel or perpendicular to the c axis is certainly associated with the evolution of the FS at $P_c$. The FS reconstruction switches from a FS analog to LaRh$_2$Si$_2$ with a localized behavior of the 4f electron ($P<P_c$) to, in the PM phase ($P>P_c$), a FS where the 4f electrons are considered to be itinerant \cite{Ara:01}.

A simple conservative picture to understand the appearance of a broad pressure slot close to $P_c$ is to assume a disappearance of AF via a first order transition at $P_c$ around 1.15 GPa, but there is an inherent pressure gradient $\nabla P$ (around 0.005 GPa) inside the pressure cell. This has been recently evaluated for example in two independant calorimetric and NMR experiments realized very carefully on the CeRhIn$_5$ system \cite{Kit:07, Bou:07}. As this inhomogeneity is quite comparable to the pressure width where a broadening of the AF phase transition appears (see the inser of figure \ref{change_V_1}) or to the pressure window where large changes occur in $\frac{\partial{T_{N1}}}{\partial{P}}$ and also $\frac{\partial{\gamma}}{\partial{P}}$, it is possible to transit from a deep broadening regime (1.155 GPa) to a regime of surviving first order transition (1.17 and 1.185 GPa).

Outside of this first experimental view, an open appealing possibility is that the collapse from AF to PM is not characterized by an unique pressure but a pressure slot \cite{Flou:06}. This P interval may be an intrinsic property governed by quantum effects without hysteresis phenomena. It is worthwhile to mention that different temperature cyclings lead to the same signal at P=1.17 GPa.

The question of a pressure spreading between the onset of first order transition at $P_c^{-}$ and its disappearance at $P_c^{+}$ was discussed for a structural transition but as emphasized spin matter offers a large diversity of situations \cite{Flou:06}. Departures from the universality of second order quantum phase transition have already been pointed out for example in MnSi \cite{Pflei:01, Doi:03, Uem:07}. where a phase separation exist over an extended P range (1.2-1.6 GPa) as well as large departures from FL behavior at least up to 2 GPa and down $T\sim$ 0.5 K.

An interesting property of the complex critical pressure regime has already been pointed out in elastic neutron scattering experiments \cite{Kawa:00}. Up to 1.03 GPa, the normalized sublatice moment, $M_0$ is proportionnal to $T_{N1}$, while at 1.08 GPa, $M_0$ decreases far more strongly than $T_{N1}$. This particular behavior supports the image of a phase separation : basically, it indicates that the fraction of AF phase decreases for P above 1.04 GPa while $T_{N1}$(P) becomes weakly P dependant. In a conventional picture, $T_{N1}$ will even be P invariant. The P change in the AF fraction will govern the apparent P collapse of the sublattice magnetization. Of course the next step now will be to combine strain gage and neutron scattering experiments in order to clarify the problem of small surviving magnetic moments above 1.04 GPa.

We have not yet studied the superconducting properties of CeRh$_{2}$Si$_{2}$ which seems related to the AF-PM transition but very sensitive to the sample's quality \cite{Ara:02, Mov:96, Kob:00}. Resistivity measurements with the current along c axis points out a full superconductivity drop only in a very narrow P window ($\Delta P = 0.05 GPa$ at P=1.05 GPa) \cite{Bou:06}. The consideration of a phase separation and associated change in spin and charge dynamics may lead to a quite furtive superconductivity. Simultaneous measurements of calorimetry and strain gage can certainly elucidate this complex quantum puzzle.

Among the interesting first order transitions in Ce HFC, a particular attention has been given to the CeCoIn$_5$ compound where magnetic quantum criticality has been pointed out just at the superconducting upper critical field H$_{c2}$(0) with magnetic field as tuning parameter despite the fact that a first order transition at $H_{c2}(T)$ appears below T=0.8K \cite{Pag:03, Bia:03, Tay:02}. In relevance to the present CeRh$_{2}$Si$_{2}$ result, an interesting observation is also a quite different behavior measured for the current parallel or perpendicular to the c axis \cite{Bia:03}. However the striking point is that the difference is now in the temperature dependence of the resistivity with close to $H_{c2}$ : $n\sim2 $ for i$\sslash$a, and $n\simeq 1$ for i$\sslash$c at a fixed orientation of H along the c axis \cite{Tan:07}. This observation coupled with the incomplet recovery of the Wiedeman law for i$\sslash$c down to 0.07 K leads to the heuristic claim of an anisotropic destruction of the FS. A new generation of experiments down to the millikelvin range is necessary to confirm or infirm the hypothesis of the Wiedmann Franz law's breakdown. Discussions on the power law need also to precise the possibility of a collisionless motion of the electrons along their orbits. In this high quality crystal ($\rho_0 \sim 0.1 \mu\Omega.cm$), such a regime is rapidly reached for moderate applied magnetic fields ($H \sim 1T$) i.e. below $H_{c2}$.

\section{Conclusions}

CeRh$_{2}$Si$_{2}$ is an excellent reference system to clarify the link between the Fermi surface (FS) evolution and heavy fermions properties under pressure. If the AF-PM phase transition is first order, a change in the localisation of the 4f particle can be easily accepted. A key problem is the FS evolution first at P=0 in the polarized paramagnetic state (PPM) state created above the metamagnetic field (H$_c\sim$23T) and then inside this PPM state under pressure, notably for $P>P_c$ \cite{Ham:00}. By comparison to YbRh$_2$Si$_2$ where similar questions are under debate, CeRh$_{2}$Si$_{2}$ is a quite more realistic experimental case since FS determinations have already been achieved inside the AF and PM phases at low field \cite{Ara:01}. Unfortunately, this situation is unlikely to happen in YbRh$_2$Si$_2$ due to the hudge value of the effective masses and the low value of the magnetic critical field ($H\simeq 0.1T$) \cite{Bou:07, Cus:03, Kneb:06}. By comparison to the CeRu$_{2}$Si$_{2}$ case where contreversies still exist on the localisation of the 4f electrons in the PPM state above H$_c$ due to the lack of a complete determination of the FS \cite{Flou:06}, the favorable ingredient of CeRh$_{2}$Si$_{2}$ is the weak value of the effective masses (5 times lower) at the metamagnetic field. That must give the opportunity of a full detection of the FS, results which are far to be achieved for CeRu$_{2}$Si$_{2}$.

\section*{Acknowledgments}
This work has been made with the ANR ICENET support.\\
Preliminary measurements on the thermal expansion of CeRh$_{2}$Si$_{2}$ were realized in Nagoya thanks to Pr. N. K. Sato (see \cite{Bou:06}).

\end{document}